# Wavelength dependence of laser-induced nanowelded microstructures assembled from metal nanoparticles

Ariel Rogers[a], Isabelle I. Niyonshuti[b], Jun Ou[c,d], Diksha Shrestha[a,e], Jingyi Chen[b,f], Yong Wang[a,e,f,*]

[a]Department of Physics, University of Arkansas, Fayetteville, Arkansas, USA, 72701.
[b]Department of Chemistry and Biochemistry, University of Arkansas, Fayetteville, Arkansas, USA, 72701.
[c]School of Engineering, California State Polytechnic University Humboldt, Arcata, California, USA, 95521.
[d]Mechanical Engineering Program, California State Polytechnic University Humboldt, Arcata, California, USA, 95521.
[e]Cell and Molecular Graduate Program, University of Arkansas, Fayetteville, Arkansas, USA, 72701.
[f]Materials Science and Engineering Program, University of Arkansas, Fayetteville, Arkansas, USA, 72701.
[*]Corresponding author. Email address: yongwang@uark.edu (YW).

## Abstract

Light-based nanowelding of metallic nanoparticles is of particular interest because it provides convenient and controlled means for the conversion of nanoparticles into microstructures and fabrication of nanodevices. Here, we demonstrated the wavelength dependence of laser-induced nanowelded shapes of silver nanoparticles (AgNPs). We observed that the nanowelded microstructures illuminated by the 405 nm laser only were more branched than those formed with illumination of both the 405 nm and 532 nm lasers. We quantified this observation by several compactness descriptors and examined the dependence of the power of the 532 nm laser. More importantly, to understand the experimental observations, we formulated and tested a hypothesis by calculating the wavelength-dependent electric field enhancement due to surface plasmon resonance of the AgNPs and nanowelded microstructures when illuminated with lights at the two wavelengths. Based on the different patterns of "hot spots" for welding AgNPs from these calculations, numerical simulations successfully "reproduced" the different shapes of nanowelded microstructures, supporting our hypothesis. This work suggests the possibility of light-based controlling the shapes of laser-induced nanowelded microstructures of metallic nanoparticles. This work is expected to facilitate the development of better nanowelding strategies of metallic nanoparticles for broader applications.

# Introduction

Due to the broad applications of metallic nanoparticles in photonics, electronics, and the fabrication of nanodevices [1–9], welding between nanoparticles (termed nanowelding) has been of great interest [10–19]. For example, nanowelding is an attractive strategy to significantly reduce the contact resistance between nanoparticles due to the capping layer of the nanoparticles during synthesis, creating solid nanocontact for efficient electron transport and facilitating the fabrication of effective nanodevices [10–14]. Nanowelding of nanomaterials has been achieved through various methods, including thermal heating, pressure, laser/light, electric current/field, and plasma [15,20–23]. Among these methods, the laser/light-based nanowelding has been particularly interesting to many researchers and us due to its non-contact and non-invasive nature, the convenience and ease of use, the wide range of choices of light wavelength, and the capability of nanowelding at room-temperature [10–12,16,24–28]. For instance, shining lasers or fluorescent lights on metallic nanoparticles could nanoweld them into higher-order assemblies and microstructures, or change their shapes in a controlled manner [11,24–27].

The light-based nanowelding relies on the localized surface plasmon resonance (SPR) in metallic nanoparticles [10,12,24–26]. Because SPR is due to the interaction of conduction electrons in the nanoparticles with the photons of the incident light and dependent on the size, shape, and composition of the nanoparticles, lights at appropriate wavelengths (i.e., around the resonance wavelength) are required to generate localized SPR for nanowelding for a given type of metallic nanoparticles [24,25,29,30]. Therefore, a single wavelength of light (or a narrow bandwidth) was typically used for the light-based nanowelding in the past, while the wavelength dependence of nanowelding of metallic nanoparticles has been rarely studied [10,15,28,29,31,32].

We set out to investigate how the shapes of nanowelded microstructures of metallic nanoparticles depend on the wavelength of light. The rationale of this work is two-fold. First, an advantage of light-based nanowelding lies in the wide range of wavelength of light [29], while understanding the wavelength dependence of laser-induced nanowelded shapes of metallic nanoparticles is expected to allow us to make use of this advantage. Second, although single metallic nanoparticles (of given shape, size, and composition) has a single, distinct resonance wavelength, nanowelded assemblies and microstructures have different shapes and sizes, and thus are likely to show different SPR responses to different wavelengths [24,25,29].

In this study, we used silver nanoparticles (AgNPs) as a model and investigated the wavelength dependence of laser-induced nanowelded shapes of AgNPs. AgNPs were used in our previous study for examining the real-time kinetics of laser-induced nanowelding, in which we observed that AgNPs illuminated by a 405 nm laser formed branched microstructures [12]. Here, we illuminated the AgNPs with an additional laser at 532 nm, and observed the shapes of nanowelded microstructures using fluorescence microscopy. We also varied the power of the 532 nm laser, and quantified how the nanowelded shapes depend on the power. More importantly, to understand the experimental observations, we formulated and tested a hypothesis by calculating the wavelength-dependent electric field enhancement due to surface plasmon resonance of the AgNPs and nanowelded microstructures when illuminated with lights at the two wavelengths.

# Materials and Methods

## Synthesis and characterization of AgNPs

AgNPs used in this study were synthesized by the polyol method [33], following the same procedures as in our previous studies [12,34–36]. Briefly, 50 mL of ethylene glycol (EG, J.T. Baker) was added to a 250-mL 3-neck round bottom flask and heated to 150 °C in an oil bath, followed by adding 0.6 mL of 3 mM NaHS (Alfa Aesar) in EG, 5 mL of 3 mM HCl (Alfa Aesar) in EG, 12 mL of 0.25 g polyvinylpyrrolidone (PVP, MW ~55,000, Sigma-Aldrich) in EG, and 4 mL of 282 mM $CH_3COOAg$ (Alfa Aesar). The reaction proceeded at 150 °C for ~1 h until the absorbance peak position of reaction mixture reached ~420 nm. The reaction was then quenched by placing the flask in an ice bath. Acetone was added to the mixture at a 5:1 volume ratio, and the product was collected by centrifugation. The resultant AgNPs were purified using water, collected by centrifugation, and re-suspended in water for characterization and future use, while the concentration was determined by inductively coupled plasma mass spectrometry (ICP-MS) (Thermo Scientific iCap Quadrupole mass spectrometer).

The synthesized AgNPs were characterized by transmission electron microscopy (TEM) and UV-Vis spectrometry. TEM images were captured using a TEM microscope (JEOL JEM-1011) with an accelerating voltage of 100 kV. Particle size and shape were measured on the TEM images using ImageJ and an algorithmic analysis reported by Laramy et. al. [37]. UV-Vis spectra were obtained using a UV-Vis spectrophotometer (Agilent Cary 50).

## Fluorescence microscopy for imaging of nanowelded microstructures of AgNPs

Prepared AgNPs were first diluted in ultrapure water (≥ 17.5 $M\Omega$ ·cm) to a final concentration of 1.3 mg/mL and then added to a flow chamber made of cleaned coverslips and glass slides [12,35,36,38,39]. The flow chamber was sealed using nail polish and then mounted on a fluorescence microscope for imaging [38]. The fluorescence microscope used in this study was an Olympus IX-73 inverted microscope equipped with an Olympus 100× N.A.=1.49 oil immersion TIRF objective, a multi-color laser bank (iChrome MLE, Toptica Photonics AG) and an EMCCD camera (Andor, MA), similar to our previous study [12]. The 405 nm and 532 nm lasers from the laser bank were used in this study. The microscope and data acquisition were controlled by Micro-Manager [40,41]. A BrightLine full-multiband 1$\lambda$ P-V RWE super-resolution laser filter set (LF405/488/532/635-B-000, Semrock, Lake Forest, IL) was used in the emission path of the microscope. The laser-induced nanowelded microstructures of AgNPs were imaged on the EMCCD camera. The effective pixel size of acquired images was 160 nm. To facilitate comparisons among measurements under different conditions, movies of 6000 frames were taken. The lasers was turned on at time 0 (i.e., frame 0 of the recorded movies), while the last frame of each movie was used for further analysis. The total exposure time to lasers was 5.25 min.

## Automatic identification of nanowelded microstructures of AgNPs

The laser-induced nanowelded microstructures were identified automatically using custom Python scripts based on the *scikit-image* Python package [42], as described in our previous work [12]. For the last frame of the movies, the background was first removed using a rolling-ball algorithm [43] with a ball size of 9 pixels, followed by smoothing twice using a Gaussian filter with a standard deviation of 1 pixel. The background in the smoothed image was removed once again, followed by applying a threshold to obtain a black/white (BW) image. Edges were detected from the BW image using the Sobel filter [44], followed by dilating the edges by 3 pixels to fill possible gaps in the edges. Small objects with areas <32 pixels were removed, before performing a flood fill [45]. The filled objects were eroded with 4 pixels, followed by removing small objects (area <32 pixels). The resulting BW image was segmented into individual structures, which corresponded to the identified microstructures. The boundaries of the microstructures were recorded as $n \times 2$ arrays, $B = \begin{pmatrix} x_1 & x_2 & \cdots & x_n \\ y_1 & y_2 & \cdots & y_n \end{pmatrix}^T$

## Shape analysis of the nanowelded microstructures of AgNPs

The shapes of the nanowelded microstructures of AgNPs were analyzed and compared by three compactness descriptors [46]. The first compactness descriptor is the coverage coefficient $\Omega$, defined as the covering percentage of the microstructure over its circumscribed circle. In more details, for a given microstructure described by its boundary $B$, the area of the microstructure $A$ was calculated by the Gauss's area formula [47]. The radius of the circumscribed circle of the microstructure was determined by $r_{max} = \max_i\{\sqrt{(x_i - x_c)^2 + (y_i - y_c)^2}\}$, where $x_c = \sum_{i=1}^n x_i /n$ and $y_c = \sum_{i=1}^n y_i /n$ are the position of the center of mass of the microstructure. Then the coverage coefficient was estimated as $\Omega = \frac{A}{\pi r_{max}^2}$. The second compactness descriptor is the area-to-perimeter ratio, $\gamma_{AP} = A/P$, where $P = \sum_{i=1}^n \sqrt{(x_i - x_{i-1})^2 + (y_i - y_{i-1})^2}$ is the perimeter of the boundary of the microstructure. The third compactness descriptor is the roughness coefficient $\sigma$ [46,48]. For a given microstructure with boundary $B$, we calculated distance of vertices to the center of mass, $r_i = \sqrt{(x_i - x_c)^2 + (y_i - y_c)^2}$, and estimated the standard deviation, $\sigma_r$. To facilitate the comparison among microstructures of different sizes, the roughness coefficient was normalized by the perimeter, $\sigma = \sigma_r/P$.

## Calculation of electric field enhancement

The electric field enhancement on the surfaces of a single AgNP and a cross-shaped microstructure (AgCS) in water was calculated using the MNPBEM17 toolbox [49–51]. The refractive indices of bulk silver from the toolbox was used, while the refractive index of water was taken to be 1.33. Plane electromagnetic waves of linear polarization were used for excitation, and the electric field enhancement was computed by summing the incoming and induced electric field [49–51]. The polarization was either in $x$ or $y$ direction, while the wavelength was either 405 nm or 532 nm. For each wavelength, the total electric fields were calculated by averaging the electric field enhancement in both polarization directions. When the AgNP or AgCS was illuminated by both lasers, the electric field enhancement from both wavelengths were added up.

## Simulation of growth of microstructures

Numerical simulations were run to examine the shapes of the microstructures when different locations of the microstructures (faces *vs.* multifaceted grooves) showed different probabilities of welding additional AgNPs. Briefly, an empty square grid (30 × 30) was prepared with a single

AgNP placed at the center as the seed. Then, additional 20 AgNPs were welded sequentially one by one, following three rules: (1) each new AgNP can only be placed next to the microstructure (i.e., the new location for welding must have a neighboring occupied site); (2) the probability of welding the new AgNP to a multifaceted-groove location (i.e., with multiple neighboring occupied sites) is $P_m$ and the probability of placing it to a microstructure face (i.e., with a single neighboring occupied site) is $1 - P_m$; (3) if multiple candidate sites are possible, a random candidate site is chosen for welding the new AgNP. For each value of $P_m$, 1000 simulations were run, generating 1000 different shapes. The shapes were analyzed as described above. We also varied $P_m$, from 0.05 to 0.95, and compared the compactness descriptors (e.g., coverage $\Omega$) for different $P_m$.

## Results and Discussions

The AgNPs used in this study were characterized as described previously and shared the same characteristics as those in our previous study [12]. TEM images showed that the AgNPs were mostly nanocubes, while a small fraction (~23%) appeared as nanospheres and nanotriangles (Fig. 1B and 1C). The average size of the AgNPs was measured to be 30 nm [12]. The absorption spectrum of the AgNPs was measured by UV–Vis spectroscopy, showing a peak at 420 nm [12].

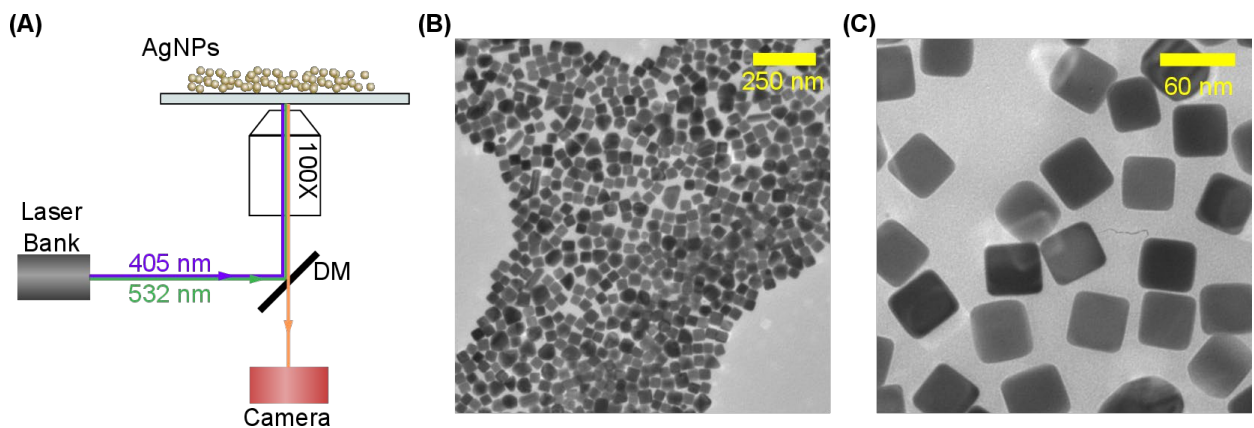

Figure 1: Experimental setup and TEM images of AgNPs. (**A**) Sketch of the experimental setup for visualizing nanowelded microstructures using fluorescence microscopy. (**B**, **C**) Representative TEM images of AgNPs used in this study.

We visualized the nanowelded microstructures formed by the AgNPs after being illuminated by a 405 nm laser, in the absence or presence of a 532 nm laser at different powers. Representative images of the nanowelded microstructures were shown in Fig. 2. In the absence of the 532 nm light, the nanowelded microstructures were branched structures (Fig. 2A and 2G), consistent with our previous report [12]. However, when the AgNPs were illuminated by both the 405 nm and 532 nm lasers, the nanowelded microstructures became more compact and less branched as the power of the 532 nm laser increased from 0% (0 mW) to 100% (~36.3 mW) (Fig. 2B–2F and 2H–2L).

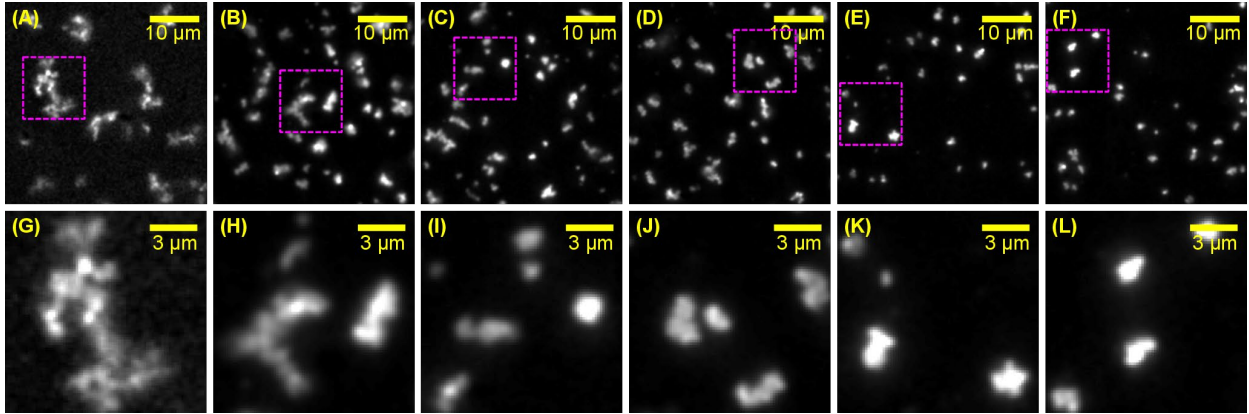

Figure 2: Representative images of nanowelded microstructures of AgNPs when illuminated with 405 nm laser of fixed power at ~3.60 mW in combination with 532 nm laser of different powers: (**A**, **G**) 0% (0 mW), (**B**, **H**) 20% (~6.00 mW), (**C**, **I**) 40% (~12.85 mW), (**D**, **J**) 60% (~20.00 mW), (**E**, **K**) 80% (~28.2 mW), and (**F**, **L**) 100% (~36.3 mW). Images in the bottom row (**G**–**L**) are zoom-in of the magenta, square regions of the top row images (**A**–**F**).

To quantify this observation, we identified the nanowelded microstructures from the images and examined the compactness of the microstructures by three compactness descriptors [46]. The first descriptor is the coverage coefficient $\Omega$, defined as the covering percentage of the microstructure over its circumscribed circle (Fig. 3A). A branched structure is expected to have a lower coverage coefficient compared to a compacter structure. As shown in Fig. 3A, the coverage coefficient of nanowelded microstructures steadily increased from 49% to 65% as the power of the 532 nm laser increased from 0% to 100%. The second compactness descriptor is the ratio of the area to the perimeter of the nanowelded microstructures, $\gamma_{AP}$. A clear increase was observed for the area-to-perimeter ratio of the nanowelded microstructures in the presence of 532 nm laser, while little dependence on the power of the 532 nm laser was seen (Fig. 3B). The third descriptor is the roughness coefficient, $\sigma$, defined as the standard deviation of the distance of boundary points to the center of the microstructure normalized by the perimeter of the microstructure [46,48]. As expected, the roughness coefficient decreased from 0.031 to 0.019 as the power of the 532 nm laser increased from 0% to 100% (Fig. 3C).

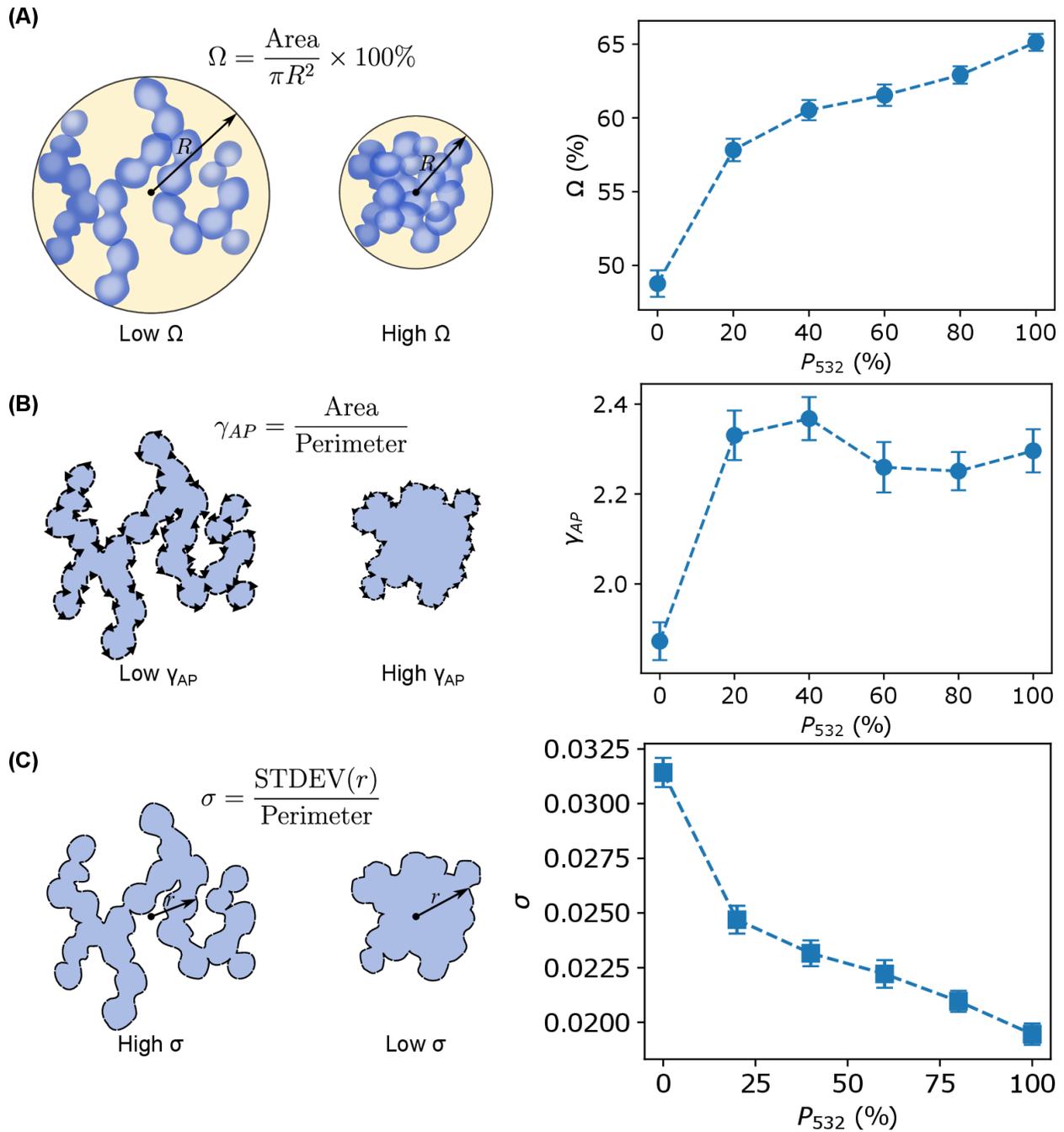

Figure 3: Compactness analysis of the nanowelded microstructures of AgNPs. (**A**) Dependence of the coverage $\Omega$ of the nanowelded microstructures on the power of the 532 nm laser. (**B**) Dependence of the area-to-perimeter ratio $\gamma_{AP}$ of the nanowelded microstructures on the power of the 532 nm laser. (**C**) Dependence of the roughness coefficient $\sigma$ of the nanowelded microstructures on the power of the 532 nm laser.

To understand the observed different nanowelded shapes of AgNPs without and with the 532 nm laser (in addition to the 405 nm laser), we hypothesized that, when illuminated by lights of different wavelengths, the surface plasmon resonance of the AgNPs and nanowelded microstructures produced different patterns of "hot spots" for the nanowelding of additional AgNPs. To test this hypothesis, we calculated the enhancement of electric field at the surfaces of a single AgNP and a cross-shaped microstructure (AgCS), using a boundary element method (BEM) approach with the MNPBEM17 toolbox [49–51]. The AgNP was modeled as a 30 × 30 × 30 nm nanocube (Fig. 4A), while the AgCS consisted of five nanocubes, with a dimension of 90 × 90 × 30 nm (Fig. 4B). The scattering spectra of the AgNP and the AgCS were calculated using the MNPBEM17 toolbox [49–51], showing a peak of 431 nm for the AgNP (Fig. 4C, black solid line), consistent with our previous result based on the discrete dipole approximation (DDA) method using the DDSCAT 7.3 program [12,52,53]. A blue shift was observed for the AgCS (Fig. 4C, red dotted line), showing a peak at 419 nm. In addition, compared to the single AgNP, the AgCS showed higher scattering at longer wavelength (Fig. 4C).

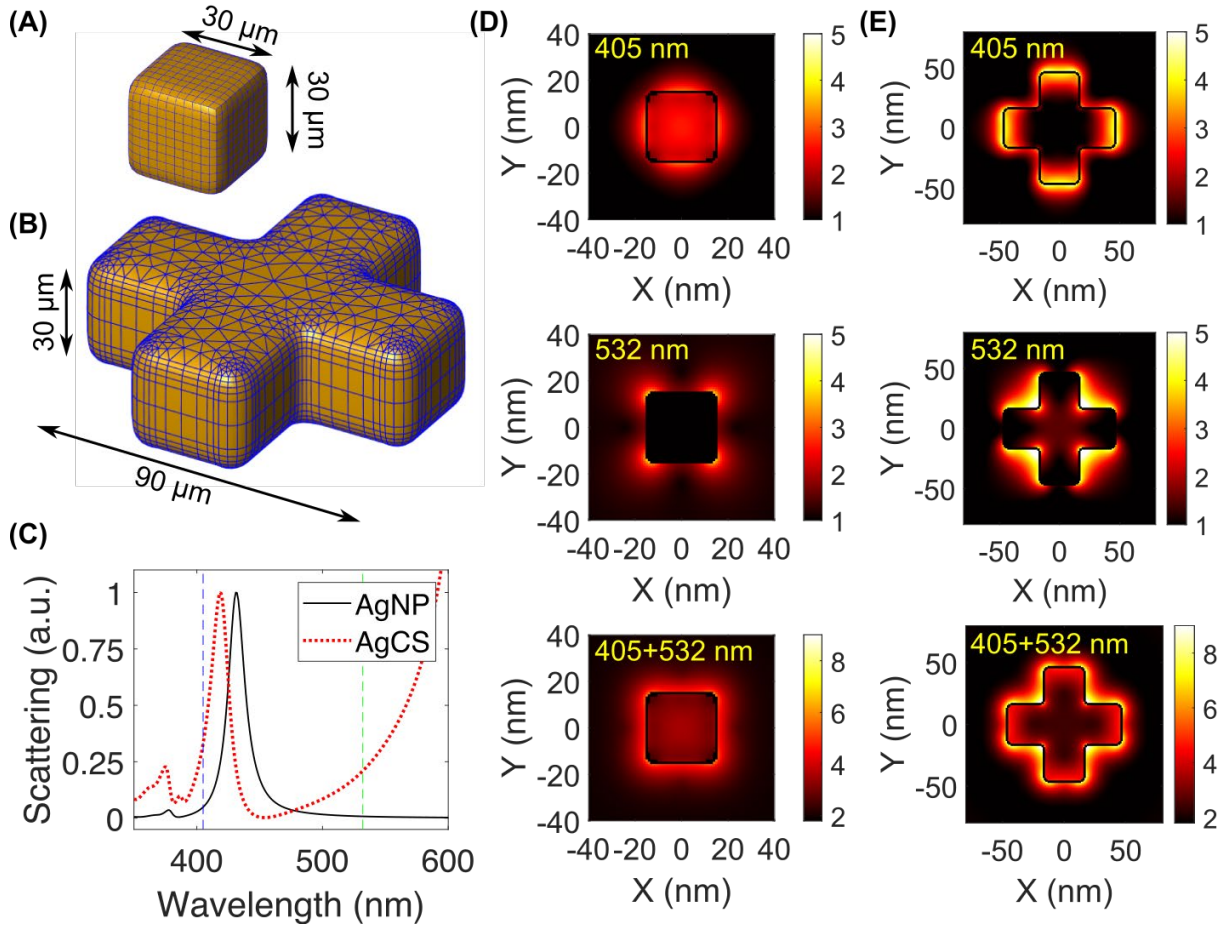

Figure 4: Electric field enhancement due to surface plasmon resonance at the surfaces of a single AgNP and a nanowelded cross-shaped microstructure. (**A**, **B**) Geometry of (**A**) a single AgNP and (**B**) a nanowelded cross-shaped microstructure (AgCS). Blue lines indicate the edges of the boundary elements generated by the MNPBEM17 toolbox [49–51]. (**C**) Calculated scattering spectra of the single AgNP and the cross-shaped AgCS. (**D**) Calculated electric field enhancement of the single AgNP when illuminated by (top) 405 nm laser only, (middle) 532 nm laser only, and (bottom) 405 and 532 nm lasers. (**E**) Calculated electric field enhancement of the cross-shaped AgCS when illuminated by (top) 405 nm laser only, (middle) 532 nm laser only,

and (bottom) 405 and 532 nm lasers.

We further calculated the electric field enhancement at the surfaces of the AgNP and the AgCS when illuminated by the 405 nm laser and/or the 532 nm laser using the MNPBEM17 toolbox [49–51]. Linearly polarized plane waves (in $x$ or $y$ direction) with wavelength of 405 nm or 532 nm were used. The electric field enhancement (averaged over the two polarization directions) of the AgNP and the AgCS at the Z=0 nm planes were shown in Fig. 4D and 4E, respectively. We observed that the electric field enhancement was more significant at the faces of the AgNPs with the 405 nm illumination, while it was more significant at the edges/corners with the 532 nm illumination (Fig. 4D). With both lasers of equal amplitude, the sum of the electric field enhancement around the AgNP was roughly uniform (Fig. 4D). The observed difference in the distribution of electric field enhancement with different illuminations was more obvious for the AgCS (Fig. 4E). With the 405 nm illumination, the electric field enhancement was significant at the faces of the AgCS. In contrast, the grooves (concave corners) of the AgCS showed the highest electric field enhancement with the 532 nm illumination (Fig. 4E). In addition, with both lasers of each amplitude, the multifaceted grooves of the AgCS remained to show the highest electric field enhancement (Fig. 4E).

To further test our hypothesis, we developed a model and numerically simulated the nanowelding growth of microstructures based on the different patterns of electric field enhancement with different illuminations. Because the generated heat is proportional to the square of the electric field ($Q \sim \sigma_{Ag} E^2$, where $\sigma_{Ag}$ is the conductivity of silver and $E$ is the electric field) [2], different patterns of electric field enhancement would result in different patterns of temperature enhancement (i.e., "hot spots") at the surfaces of the AgNPs and nanowelded microstructures. Thus, the hot spots would lead to different melting patterns and probabilities to weld new/additional AgNPs. Based on the calculations above, we therefore modeled that illumination with a single laser at 405 nm leads to a higher probability to attach new AgNPs to the faces of the microstructures (and lower probability to grow at the multifaceted grooves – low $P_m$), while illumination with both lasers (405 nm and 532 nm) have a higher probability to weld new AgNPs to the multifaceted grooves of the microstructures (high $P_m$, Fig. 5A). 1000 simulations were run for each case (i.e., 405 nm only *vs.* 405 nm + 532 nm), while the final size of microstructures was set to 21. Representative images of simulated microstructures were shown in Fig. 5B, where the top row was with $P_m = 0.1$ (i.e., illumination with 405 nm only) and the bottom row was with $P_m = 0.9$ (i.e., illumination with 405 nm + 532 nm). We observed that illumination with a single laser at 405 nm resulted in branched structures (top row of Fig. 5B), while illumination with both lasers led to compact structures (bottom row of Fig. 5B), consistent with our experimental results (Fig. 2). Furthermore, we quantified the compactness of the simulated microstructures in each case by the coverage coefficient $\Omega$ and observed that the coverage coefficient increased as the increased (Fig. 5C), again consistent with the experimental results (Fig. 3).

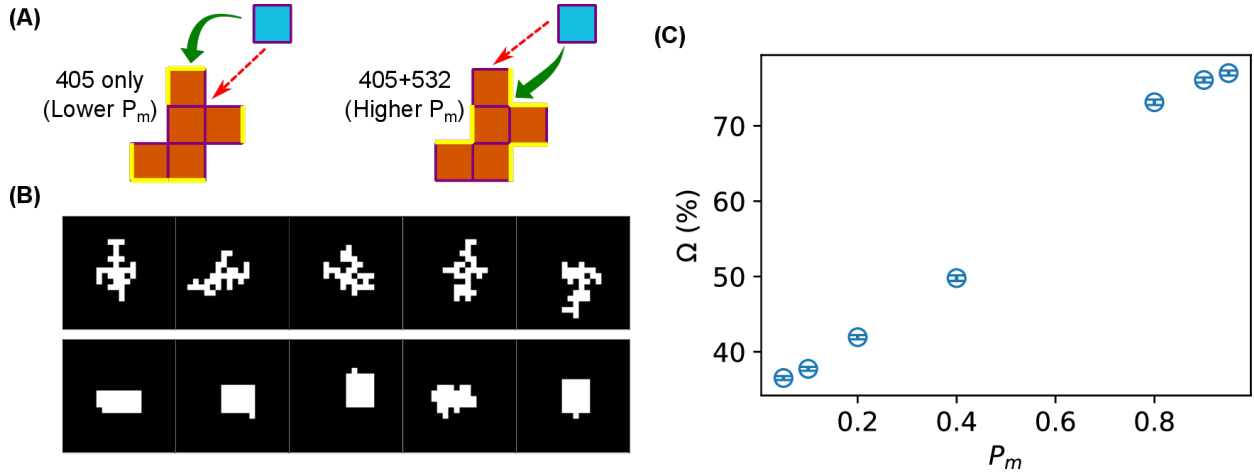

Figure 5: Simulation of different shapes of nanowelded microstructures based on the differences in the SPR-induced electric field enhancement. (**A**) Model for the growth of nanowelded microstructures with different probabilities of growing at the corners (i.e., newly attached AgNP is adjacent to multiple faces), $P_m$. The yellow edges highlight the preferences of attachment of new AgNPs with low (left) or high (right). (**B**) Representative shapes of nanowelded microstructures with (top) $P_m = 0.1$ and (bottom) $P_m = 0.9$. (**C**) Dependence of the coverage $\Omega$ of the simulated nanowelded microstructures on the $P_m$.

## Conclusions

In conclusion, we investigated the wavelength dependence of laser-induced nanowelded shapes of AgNPs. We observed that the nanowelded microstructures of AgNPs illuminated by the 405 nm laser only were more branched than those formed with illumination of both the 405 nm and 532 nm lasers (Fig. 2). We quantified this observation by several compactness descriptors and examined the dependence of the power of the additional 532 nm laser (Fig. 3). More importantly, to understand the experimental observations, we formulated and tested a hypothesis by calculating the wavelength-dependent electric field enhancement due to surface plasmon resonance of the AgNPs and nanowelded microstructures when illuminated with lights at the two wavelengths (Fig. 4). Based on the different patterns of "hot spots" for welding additional AgNPs from these calculations, we ran numerical simulations and successfully "reproduced" different shapes of nanowelded microstructures (Fig. 5), supporting our hypothesis.

Fluorescence imaging has been used in our studies to examine the nanowelded microstructures of AgNPs [12], which was possible due to the photoluminescence of AgNPs [35,54,55]. Compared to electron microscopy, fluorescence microscopy is simple, versatile, and convenient. As other metallic nanoparticles (e.g., gold and copper nanoparticles) are also photoluminescent [56–61], we expect that methods exploited in the current study are readily applicable to the studies of nanowelding of other metallic nanoparticles.

The current work suggests that it is possible to control the shapes of laser-induced nanowelded microstructures of metallic nanoparticles using light of different wavelengths. This capability of light-based control of nanowelded shapes is expected to be convenient for various applications, such as conductive bonding [15,62,63] and nanoparticle-based metal printing [64–66]. Our study is expected to provide a better understanding and control in such applications. We showed that the observed wavelength dependent nanowelded shapes are because of the wavelength dependence of the electric field enhancement or the surface plasmon resonance. It would be interesting to theoretically and experimentally explore the dependence of nanowelded shapes of metallic nanoparticles on a wider range of wavelengths. We expect that such work will facilitate rationale design of different shapes using illuminations of different wavelengths.

## Acknowledgment

This work was supported by the National Science Foundation (Grant No. 1826642), the National Institute of Food and Agriculture / United States Department of Agriculture (Grant No. 2021-67019-33683), and the Arkansas Biosciences Institute (Grants No. ABI-0189, ABI-0226, ABI-0277, ABI-0326, ABI-2021, and ABI-2022). We are also grateful for support from the Arkansas High Performance Computing Center (AHPCC), which is funded in part by the National Science Foundation (Grants No. 0722625, 0959124, 0963249, and 0918970) and the Arkansas Science and Technology Authority.

## Author Contributions

YW conceived and designed the project; AR performed experiments and acquired the data; IIN and JC synthesized and characterized the nanoparticles; AR, JO, DS, and YW performed data analysis and visualization; JO and YW performed simulations; AR and YW wrote the paper; all authors reviewed, commented, and revised the paper.